\documentclass[a4paper,11pt]{article}
\usepackage[utf8]{inputenc}

\usepackage{graphicx}
\usepackage{epsfig}
\usepackage[export]{adjustbox}
\usepackage{subcaption}
\usepackage{epstopdf}
\usepackage{setspace}

\usepackage{nameref,hyperref}
\usepackage[right]{lineno}

\usepackage{latexsym}
\usepackage{amsmath}
\usepackage{amssymb}
\usepackage{pifont}
\usepackage{amsfonts}
\usepackage{lmodern,bm}  
\usepackage{mathtools}

\usepackage{times}
\usepackage{xcolor}

\newcommand{\be}{\begin{equation}}
\newcommand{\ee}{\end{equation}}
\newcommand{\bea}{\begin{eqnarray}}
\newcommand{\eea}{\end{eqnarray}}
\newcommand{\bac}{\begin{array}{|c|}}
\newcommand{\eac}{\end{array}}
\newcommand{\noi}{\noindent}

\newcommand{\eg}{e.g. }
\newcommand{\ie}{i.e. }

\newcommand{\kap}{\kappa}

\begin{document}

\title{Allometric scaling laws derived from symmetric \\ tree networks}
  
\author{L. Zavala Sans\'on \\
\normalsize{Departamento de Oceanograf\1a F\1sica, CICESE, M\'exico} \\ \\
A. Gonz\'alez-Villanueva \\
\normalsize{Colegio de Ciencia y Tecnolog\1a, Plantel San Lorenzo Tezonco,} \\
\normalsize{Universidad Aut\'onoma de la Ciudad de M\'exico, M\'exico. } }

\date{\today}

\maketitle

\begin{abstract}

A set of general allometric scaling laws is derived for different systems represented by tree networks. The formulation postulates self-similar networks with an arbitrary number of branches developed in each generation, and with an inhomogeneous structure given by a fractal relation between successive generations. Three idealized examples are considered: networks of masses, electric resistors, and elastic springs, which obey a specific recurrence relation between generations. The results can be generalized to networks made with different elements obeying equivalent relations. The equivalent values of the networks (total mass, resistance and elastic coefficient) are compared with their corresponding spatial scales (length, cross-section and volume) in order to derive allometric scaling laws. Under appropriate fractal-like approximations of the length and cross-section of the branches, some allometric exponents reported in the literature are recovered (for instance, the 3/4-law of metabolism in biological organisms or the hydraulic conductivity scaling in porous networks). The formulation allows different choices of the fractal parameters, thus enabling the derivation of new power-laws not reported before. 

\end{abstract}

\newpage

\section{Introduction}

Complex networks have been thoroughly studied in the last decades due to the wide range of applications in several scientific and technological fields, such as distribution networks, biological systems, computational arrays or social networks \cite{Boccaletti06},  \cite{Albert02}. 
In this paper we focus our attention on tree-like networks, which develop their structure as sets of branches starting from a common origin. Tree networks are of special interest in several research fields because its branching structure is suitable to model biological trees, ecological networks, rivers, neural structures, atmospheric discharges, plasma lamps, among many others (see \eg \cite{Harris89}, \cite{Turcotte98}, \cite{Rinaldo98}, \cite{Franco06}). 

According to their intrinsic features, local and global properties of tree networks are valuable for understanding and, eventually, obtaining useful information for practical purposes. For instance, \cite{Franco06} investigated the control of branch flow rates and fault detection in water distribution systems. In Ref. \cite{Pelletier00} the similarity between the structure of river networks and the vein structure of leaves is reported (see also \cite{Sanchez03} where those systems are compared with the structure of marine corals). The transport properties of fractal networks has been examined in different systems by \cite{Xu06}, and in particular for laminar and turbulent flows in pipes \cite{Kou14}.

In contrast with previous studies, we present a general theoretical framework for symmetric tree-like networks, and then we derive a set of general allometric laws for networks made of different elements. This is a relevant subject in several fields (engineering, hydrology, geomorphology, biology) in which empirical relations between the properties of the network and their spatial scales are commonly found in the form of power-laws (\cite{Turcotte97}, \cite{West05}). When modeled as fractal structures, the scaling laws of many of these systems have been theoretically derived; for instance: the electric resistance in terms of the volume of the network \cite{Xu06}, the hydrological conductivity in a porous media \cite{Winter01}, and the allometric laws obeyed by several biological organisms  \cite{West05}. The model developed in this paper recovers some of these expressions and, more importantly, it predicts new scaling laws that appear as a natural consequence of the networks' symmetry.

The physical nature of the elements that constitute the networks is considered: masses, resistors and springs. Thus, we are interested in global properties such as the total mass, resistance or elastic coefficient of the whole system, respectively. It is shown that these  properties depend directly on the additive rules or recurrence relations between successive generations obeyed by the elements of the network. An important implication is that the mathematical procedures can be applied to networks made of different elements, as long as these follow analogous recurrence relations. Thus, the results are far more general than those found for the three proposed examples. 

The paper is organized as follows. In the second Section the basic definitions and the general symmetry properties of the branching networks are presented. Then we study tree-like networks of masses, resistors and springs, and examine the properties of finite and infinite networks. 
Third Section is devoted to the general derivation of scaling power-laws of the previously examined networks. 
Finally, in last Section we discuss the main results to the light of previous studies, and mention some limitations of the formulation, as well as some ideas for future research.

\section{Symmetric tree-like networks}

\subsection{Basic definitions}

\begin{figure}[h]
\includegraphics[width=1\textwidth]{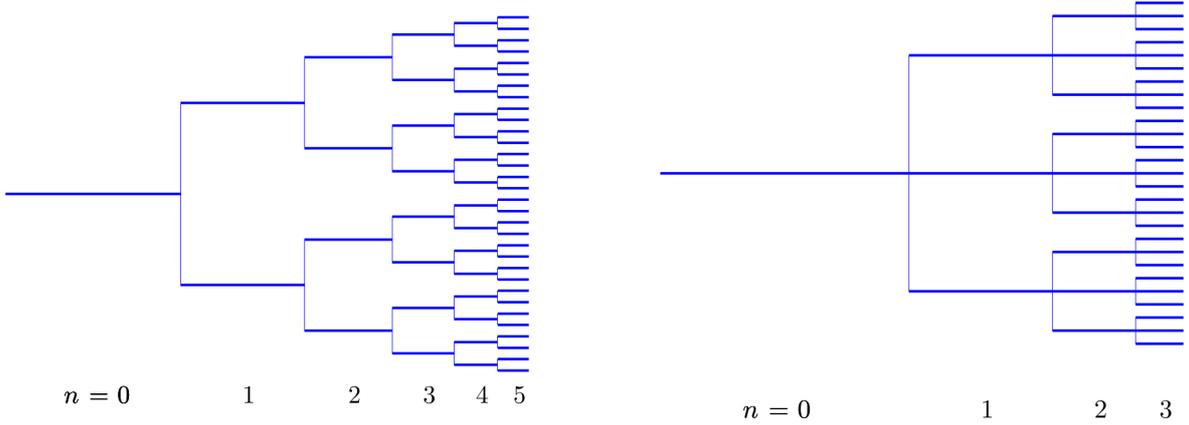}
\caption{Schematic representation of tree-like networks with branching factor $\beta$, and $n$ generations (the trunk is generation 0). Left panel: $\beta=2$, $n=5$. Right panel: $\beta=3$, $n=3$.}
\label{f1}
\end{figure}

Consider a self-similar array of {\it elements} (which can be physical objects or abstract entities) starting from a common origin, as in Figure \ref{f1}. The {\em branching factor} is a natural number $\beta\ge1$ denoting the (arbitrary) number of new elements generated from each branch \cite{Winter01}. 
A new set of branches is called a {\em generation}. The number of generations is denoted by $n$, starting from $n=0$ at the trunk. Infinite networks are allowed ($n\rightarrow \infty$). 

There is a fundamental global symmetry across the direction of growth (the horizontal axis in Figure \ref{f1}): the entire network at one side of this direction is a reflection of the other side. In addition, at any generation there are equivalent local symmetries along the same direction.  In this representation the only relevant elements are the branches in the direction of growth, \ie the perpendicular segments are drawn only for the sake of visual simplicity, with no consequences on the additive properties between generations. 

A {\it homogeneous} network is defined as an array in which the elements of all generations are identical. Thus, an {\it inhomogeneous} network is characterized by different magnitude or weight of the elements between generations. Such differences might obey a specific rule or they can be completely arbitrary. In this study we will focus on inhomogeneous networks in which successive generations are weighted by a constant factor $\alpha$ (a positive real number). For instance, for a network of resistors with electric resistance $r$ at generation zero, the elements in the first generation will have a resistance $\alpha r$, those in the second generation $\alpha^2 r$, and so on, such that $r_{j+1}=\alpha r_j$ for any generation $j$. 

The basic property to calculate is the {\it equivalent value}, which refers to the measured value when considering the total extension of the network at generation $n$. For the example of an array of resistors, the equivalent value is the total resistance of the network, from the trunk to the last generation. 

\subsection{Networks of masses, resistors and springs}

We shall consider networks whose branches are made of three types of elements: masses, resistors and springs. These physical elements were chosen because they follow different additive rules between generations. The corresponding values at the trunk ($n=0$) are denoted by $m$, $r$ and $k$, which stand for mass, electric resistance and elastic coefficient. The equivalent values of the networks after $n$ generations will be denoted with greek characters $\mu_n$, $\rho_n$ and $\kappa_n$, respectively.  

Equivalent values are determined by the corresponding additive rules of the elements. These rules lead to the following recurrence relations between generations for the three types of networks with arbitrary inhomogeneity $\alpha$ and branching degree $\beta$: 
\be
\mu_{j+1}=m+\alpha \beta \mu_{j}, \hspace{1cm} \rho_{j+1}=r+\frac{\alpha}{\beta} \rho_{j}, \hspace{1cm} \frac{1}{\kappa_{j+1}}=\frac{1}{\kappa}+\frac{1}{\alpha\beta} \frac{1}{\kappa_{j}},
\label{recrel}
\ee

\noi for any generation $j \ge 0$, and $\mu_0\equiv m$, $\rho_0\equiv r$, and $\kappa_0\equiv k$. For the network of masses the recurrence relation is evident by simply adding the mass of each branch. The recurrence relation for the network of resistors follows from the rules of parallel resistors $1/r_{eq}=\sum_i 1/r_i$ (used for branches at a given generation) and in series $r_{eq}=\sum_j r_j$ (used for successive generations). Analogoulsy, for an elastic network, the additive rules for parallel springs is simply $k_{eq}=\sum_i k_i$, whilst springs connected in series give $1/k_{eq}=\sum_j 1/k_j$. Note that the three recurrence relations have a similar form, but depend on $\alpha$ and $\beta$ in a different way. 

Applying successively the corresponding relations (\ref{recrel}), it is found that the equivalent values of mass, resistance and elastic coefficient after $n$ generations are given by the following geometrical series:
\be
\mu_n=\frac{1-(\alpha \beta)^{n+1}}{1-\alpha \beta}m, \hspace{1cm} \alpha \beta\neq 1,
\label{mrk1}
\ee

\be
\rho_n=\frac{1-(\frac{\alpha}{\beta})^{n+1}}{\kern-9pt 1-\frac{\alpha}{\beta}}r, \hspace{1cm} \frac{\alpha}{\beta}\neq 1,
\label{mrk2}
\ee

\be
\kap_n=\frac{\kern-9pt 1-\frac{1}{\alpha \beta}}{1-(\frac{1}{\alpha \beta})^{n+1}}k, \hspace{1cm} (\alpha \beta)^{-1} \neq 1.
\label{mrk3}
\ee

\noi The equivalent values given by (\ref{mrk1}), (\ref{mrk2}) and (\ref{mrk3}) are generalizations of similar expressions that have been reported in the context of  transport properties in different physical systems (\cite{Franco06},  \cite{Xu06}),  \cite{Kou14}). The relevance of the present formulation is that one can examine these properties for an arbitrary number of branches $\beta$ and inhomogeneity $\alpha$, for both finite and infinite arrangements. 
In a preliminary study by the authors \cite{Gonzalez15} the equivalent values for resistors (\ref{mrk2}) and springs (\ref{mrk3}) are plotted against $\alpha / \beta$ and  $(\alpha \beta)^{-1}$, respectively, for different $n$ values.
A remarkable property when $n\rightarrow \infty$ is that the equivalent values might be finite depending on the product or ratio between $\alpha$ and $\beta$. 
For instance, for large $\alpha$ (say 10) the total resistance of the infinite network will remain finite as long as it generates $\beta=11$ or more branches, such that $\alpha/\beta<1$. This property was noticed by \cite{Doyle00} for a bifurcating ($\beta=2$), homogenous ($\alpha=1$) electric network.

\subsection{Spatial scales}

Consider now the length, cross-section and volume of the networks. For this purpose, it is assumed that the spatial scales of the successive generations grow in a fractal-like fashion: (i) The length of the branches changes by a  factor $\gamma$ each generation, so $\ell_{j+1}=\gamma \ell_j$. (ii) The cross-section of the branches is scaled by a factor $\sigma$, such that  $a_{j+1}=\sigma a_j$. (iii) Consequently, the volume changes as $v_{j+1}=\gamma \sigma v_j$. The scales of the trunk (generation 0) are denoted as $\ell$, $a$ and $v$.

The total length of the network after $n$ generations is calculated as the sum of the length of single branches at each generation 
\be
L_n=\sum^n_{j=0}\gamma^j\ell=\frac{1-\gamma^{n+1}}{\kern-9pt 1-\gamma}\ell, \hspace{1cm} \gamma \neq 1.
\label{Ln}
\ee

\noi Note that this is a measure {\em along} the network. In contrast, a measure {\em across} the network is the sum of the cross-section area of the $\beta^n$ branches at generation $n$:
\be
A_n=(\beta \sigma)^n a.
\label{An}
\ee

\noi Finally, the total volume of the whole network is measured by considering the volume of all branches in all generations:
\be
V_n=\sum^n_{j=0}(\beta \gamma \sigma)^j v=\frac{1-(\beta \gamma \sigma)^{n+1}}{\kern-9pt 1-\beta \gamma \sigma}v, \hspace{1cm} \beta \gamma \sigma \neq 1.
\label{Vn}
\ee

\section{Allometric laws}

\subsection{Scaling relations}

In order to derive appropriate scaling relations for the types of networks introduced above, it is convenient to apply the following procedures.
The spatial scales of the trunk can be written in terms of the corresponding values at generation $n$: $\ell=\ell_n/\gamma^n$, $a=a_n/\sigma^n$ and $v=v_n/(\gamma \sigma)^n$, respectively. Using these expressions and defining nondimensional values $\hat L=L_n/\ell_n$, $\hat A=A_n/a_n$, $\hat V_n=V/v_n$, Eqs (\ref{Ln}) to (\ref{Vn}) can be written as 
\be
\hat L=\frac{1-\gamma^{n+1}} {(1-\gamma) \gamma^n}  ,\,\,\,\,\,0<\gamma<1,
\label{lav1} 
\ee

\be
\hat A=\beta^n,
\label{lav2}
\ee

\be
\hat V=\frac{1-(\beta \gamma \sigma)^{n+1}} {(1-\beta \gamma \sigma) (\gamma \sigma)^n},  \,\,\,\,\,0<\gamma \sigma<1.
\label{lav3}
\ee

\noi The restrictions on $\gamma$, and $\gamma \sigma$ are necessary conditions for having finite scale networks (length and volume do not diverge for large $n$). 

Equivalently, the seed values for mass, resistance and elastic coefficient can be written as $m=m_n/\alpha^n$, $r=r_n/\alpha^n$ $k=k_n/\alpha^n$, respectively. Using these expressions and defining $\hat M=\mu_n/m_n$, $\hat R=\rho_n/r_n$, $\hat K=\kappa_n/k_n$, Eqs (\ref{mrk1}), (\ref{mrk2}) and (\ref{mrk3}) yield
\be
\hat M=\frac{1-(\alpha \beta)^{n+1}}{(1-\alpha \beta)\alpha^n},\,\,\,\,\,0<\alpha \beta<1,
\label{mrk1b}
\ee

\be
\hat R=\frac{1-(\frac{\alpha}{\beta})^{n+1}}{(1-\frac{\alpha}{\beta})\alpha^n},\,\,\,\,\,0<\frac{\alpha}{\beta}<1,
\label{mrk2b}
\ee

\be
\hat K=\frac{1-\frac{1}{\alpha \beta}}{[1-(\frac{1}{\alpha \beta})^{n+1}]\alpha^n},\,\,\,\,\,0<\frac{1}{\alpha \beta}<1.
\label{mrk3b}
\ee

By taking logarithms of Eqs (\ref{lav1}) to (\ref{mrk3b}) and approximating for large $n$ (in order to neglect terms with exponent $n+1$), it is possible to obtain nine scaling laws for $\hat M$, $\hat R$, and $\hat K$ with respect to the spatial scales $\hat L$, $\hat A$, and $\hat V$. For instance, after several generations $n$
\bea
\ln \hat M &\approx& -\ln(1-\alpha\beta)-n\ln\alpha \approx  -n\ln\alpha \nonumber \\ \\
\ln \hat L &\approx& -\ln(1-\gamma)-n\ln\gamma \approx  -n\ln\gamma. \nonumber
\eea

\noi Thus, the ratio $\ln \hat M/\ln \hat L \approx \ln \alpha/\ln \gamma$ indicates that the total mass obeys a power-law with respect to the total length scale, $\hat M \approx \hat L^{\ln \alpha/\ln \gamma}$. In fact, the ratios for mass, resistance and elastic coefficient with respect to length have identical expressions, as well as those with respect to area and volume:
\be
\frac{\ln \hat M}{\ln \hat L}=\frac{\ln \hat R}{\ln \hat L}=\frac{\ln \hat K}{\ln \hat L}\approx \frac{\ln \alpha}{\ln \gamma},
\label{lnl}
\ee

\be
\frac{\ln \hat M}{\ln \hat A}=\frac{\ln \hat R}{\ln \hat A}=\frac{\ln \hat K}{\ln \hat A}\approx -\frac{\ln \alpha}{\ln \beta},
\label{lna}
\ee

\be
\frac{\ln \hat M}{\ln \hat V}=\frac{\ln \hat R}{\ln \hat V}=\frac{\ln \hat K}{\ln \hat V}\approx \frac{\ln \alpha}{\ln (\gamma \sigma)}.
\label{lnv}
\ee

In order to obtain specific values of the coefficients at the right-hand side of these expressions, it is necessary to make further assumptions regarding the inhomogeneity introduced by $\alpha$, as well as on $\sigma$ and $\gamma$. Appropriate choices of these three parameters define a network model 
that obeys power-laws with respect to length, area and volume. In what follows we consider a particular approach, but it must be kept in mind that there might be much more possibilities.

\subsection{Basic allometric model}

In order to find a suitable expression for $\sigma$, it is assumed that the sum of the cross-section of the branches in a new generation is the same as the cross section of the parent branch. In other words, the area between generations is preserved, which means that $\beta a_{j+1}=a_j$. Comparing with the fractal scaling $a_{j+1}=\sigma a_j$ it is verified that 
\be
\sigma=\beta^{-1}
\label{allo1}
\ee

\noi Note that the area preserving condition says that the cross-section of the trunk is equal to the total cross-section at any generation, $a=(\beta \sigma)^n a$, which also implies (\ref{allo1}).

For the length factor $\gamma$ we use the ``space-filling" assumption formulated in Ref. \cite{West97}, which provides a volumetric relation based on the length of the branches between successive generations, that is, $\beta \ell_{j+1}^3=\ell_j^3$.
Since $\ell_{j+1}=\gamma \ell_j$, this condition implies that
\be
\gamma=\beta^{-1/3}.
\label{allo2} 
\ee

Regarding $\alpha$, it is useful to consider the preservation of an intensive variable in all branches of the network. Such a variable is intrinsically different in the different networks, so they have to be analyzed separately.

\subsubsection*{Network of masses} 

Consider a network of masses in which every branch at any generation has the same mass density $D_m$. Let the mass of a given branch at generation $j$ be $m_j=D_m \ell_j a_j$. Then, the mass of a branch at the next generation is $m_{j+1}=D_m \ell_{j+1} a_{j+1}= \gamma \sigma m_j$. This sets the value of the inhomogeneity as $\alpha=\gamma \sigma$. Using this expression together with (\ref{allo1})  and (\ref{allo2}) yields
\be
\alpha =  \beta^{-4/3}.
\label{allo3}
\ee

\subsubsection*{Electric network} 

For the network of resistors, the equivalent assumption is the preservation of resistivity $D_r$ in every branch at any generation. Thus, if resistance at level $j$ is $r_j=D_r \ell_j/a_j$, then the resistance of a branch at the next generation is $r_{j+1}=\rho \ell_{j+1}/a_{j+1}= (\gamma/ \sigma) r_j$. The value of the inhomogeneity yields $\alpha=\gamma/\sigma$, which implies when using (\ref{allo1})  and (\ref{allo2}) again:
\be
\alpha=\beta^{2/3}.
\label{allo4}
\ee

\subsubsection*{Elastic network} 

Finally, an equivalent expression for the elastic network implies that $k_j=D_k/(\ell_j a_j)$. The coefficient of a branch at the next generation is $k_{j+1}=D_k/(\ell_{j+1}a_{j+1})= k_j/(\gamma \sigma)$. The value of the inhomogeneity yields $\alpha=1/(\gamma \sigma)$. Therefore, 
\be
\alpha=\beta^{4/3}.
\label{allo5}
\ee

Summarizing, the allometric model for each network type is defined by Eqs (\ref{allo1}) and (\ref{allo2}) for $\sigma$ and $\gamma$, respectively, and the corresponding approximation for $\alpha$, Eq (\ref{allo3}), (\ref{allo4}) or (\ref{allo5}). Using these expressions the asymptotic logarithm ratios, given by the right-hand side of (\ref{lnl}), (\ref{lna}) and (\ref{lnv}), are calculated. The result is nine power-laws for the total mass, resistance and elastic coefficient in terms of the total length, cross-section and volume of the networks, which are presented in Table \ref{tab1}. The present theory recovers some well-known exponents reported in the literature in several fields, and predicts new scaling laws, as discussed in next section.

\begin{table}[h]
\caption{Scaling laws in the allometric model: $\sigma=\beta^{-1},\gamma=\beta^{-1/3}$, $\alpha$ as indicated.}
\begin{center}
\begin{tabular} {|lclll|}
\hline
Property & Weight $\alpha$ & Length & Area & Volume \\
\hline
& & & & \\  
Mass & $\alpha=\gamma\sigma$ & $\hat M=\hat L^4$ & $\hat M=\hat A^{4/3}$ & $\hat M=\hat V$ \\
& & & & \\
Resistance & $\alpha=\gamma/\sigma$ & $\hat R=\hat L^{-2}$ & $\hat R= \hat A^{-2/3}$ & $\hat R=\hat V^{-1/2}$ \\
& & & & \\
Elastic coeff. & $\alpha=1/(\gamma\sigma)$ & $\hat K= \hat L^{-4}$  & $\hat K= \hat A^{-4/3}$ & $\hat K= \hat V^{-1}$ \\
& & & & \\
\hline
\end{tabular}
\end{center}
\label{tab1}
\end{table}

\begin{figure}
\includegraphics[width=1\textwidth]{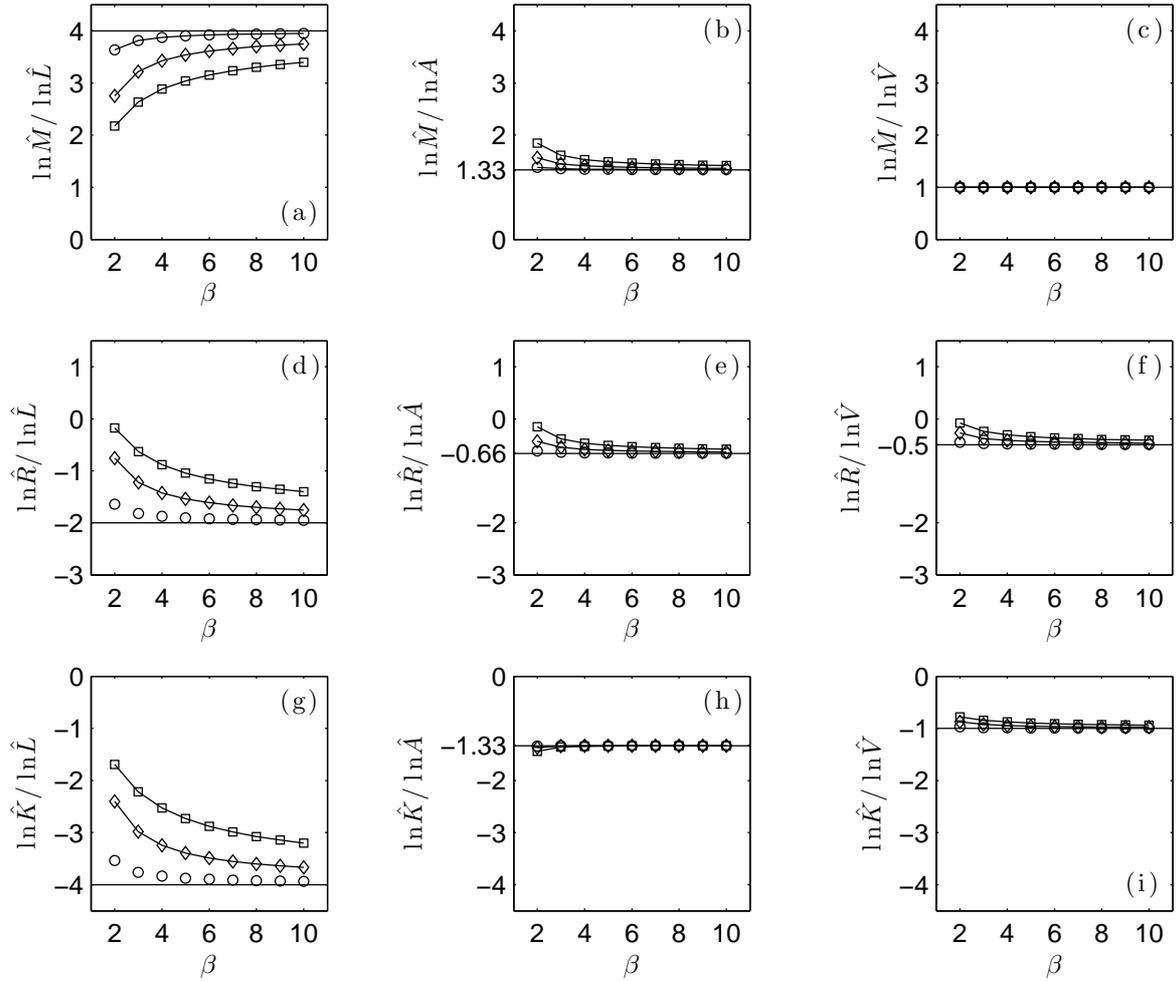}
\caption{Ratio between the logarithm of equivalent values of different networks ($\hat M, \hat R, \hat K$) and the logarithm of their spatial scales ($\hat L, \hat A, \hat V$) as a function of the number of branches $\beta$. Calculations are based on the full expression of the equivalent values, Eqs (\ref{lav1}) to (\ref{mrk3b}), for $n=3$ (squares), 9 (diamonds) and 50 (circles), and using the corresponding allometric model for $\gamma$, $\sigma$ and $\alpha$ (see text). For large $n$, the ratios tend to the predicted limits (Table \ref{tab1}) indicated by the horizontal lines.}
\label{f6}
\end{figure}

In order to test that these are the correct limits for large $n$ in the previous analyses, Figure \ref{f6} shows the logarithm ratios for different number of branches $\beta$ and generations $n$, corresponding to the three types of networks. The calculations are made by using the full expressions for $\hat M$,  $\hat R$ and  $\hat K$, as well as for  $\hat L$,  $\hat A$ and  $\hat V$ [Eqs (\ref{lav1}) to (\ref{mrk3b})], and applying the corresponding allometric model.
The first point to remark is that the asymptotic values for large $n$ are indeed those calculated previously (Table \ref{tab1}), although convergence is slower for $\beta=2$ or 3 branches. A second aspect is that the logarithm ratios with respect to the length scale (panels a, d and g) are somewhat far from the asymptotic limit when $n=3$ and 9, and low $\beta$. In contrast, the ratios with respect to area and volume converge very rapidly (panels b, c, e, f, h, i). Note that $\log \hat M/ \log \hat V=1$ for all $n$ and $\beta$ (panel c), which is an exact result because the allometric model considers branches with the same density, and therefore the total mass is proportional to the total volume.

\section{Discussion and conclusions}

We have obtained a set of allometric scaling laws associated with idealized, tree-like networks with local and global symmetries with respect to the direction of growth. 
The study is focused on networks whose basic elements are masses, resistors and springs. The scaling properties of the three types of networks with respect to length, cross-section and volume were presented in Table \ref{tab1} and in Figure \ref{f6}. The present theory recovers some scaling laws previously reported in the literature and, more importantly, predicts new scaling laws that apparently have not been studied before.

Let us mention some specific cases. There are several scaling laws in biology with exponents that are simple multiples of 1/4 \cite{West05}. In particular, a well-known empirical formula establishes that the metabolic rate $B$ of a very wide range of organisms is related with their total mass body $M_b$ with a 3/4-law, such that $B\sim M_b^{3/4}$ (see \cite{West97} and references therein). This expression is recovered in our formulation by assuming that the metabolic rate is proportional to the cross-section area associated with the network, $B \propto A$. Such an assumption, carefully justified in \cite{West99}, leads to the same scaling law for the equivalent mass in terms of the total cross-section, $\hat M\sim \hat A^{4/3}$ (Table \ref{tab1}).
In fact, the three allometric formulae for the total mass 
recover the scaling laws reported in \cite{West99}, denominated as {\it fractal biological} (see their Table 2), and which apply to different biological systems (plants, mammals and unicellular organisms). Although there has been some criticism to these scaling exponents (see \eg \cite{Dodds01}) we consider a relevant issue to find such power-laws in simplified models in order to gain a better understanding in more complex systems, as discussed in \cite{West05}.  

Another result of interest in a very different field, and reproduced with the present allometric model, is the electric resistance of a tree network in terms of the inverse of the square root of the total volume, $R\sim V^{-1/2}$, found in \cite{Xu06}  (see Table \ref{tab1}). An equivalent case is the hydraulic conductivity $C$, which can be estimated as the inverse of the total resistivity in a porous media and obeys a power-law with the total volume of the system. Using the analogy with the electric resistance one may assume $C=R^{-1}$, implying that $C\sim V^{1/2}$, which is a well-known empirical law in geohydrology \cite{Winter01}. 

Summarizing the applicability of the present theory, it is emphasized that it recovers the fractal biological scaling in a very natural way without invoking efficiency or dynamical arguments. In other words, it is less restrictive than the theory of \cite{West97} and more in the vein of \cite{West99}: the scaling laws are derived by taking advantage of the geometry of the networks, \ie their multiple symmetries. It also recovers some other power-laws for electrical and hydrological networks with no need to use elaborated hierarchichal trees \cite{Winter01}. More importantly, establishes new power-laws which would be relevant to test (specially those regarding the network of springs) and sets a solid formulation to search for different models, besides the basic allometric one used here.

Furthermore, the formulation of the networks of masses, resistors or springs can be applied to different systems. In order to emphasize this point, note that the recurrence relations (\ref{recrel}) can be applied to different elements. For instance, the additive rule for resistors is exactly the same as that for coils, whilst the corresponding rule for springs is identical to that for electric capacitors. Thus, the three cases presented here represent conceptual models of a great variety of networks which might have much more general applications. 

There are some restrictions of the model that must be taken into account, such as the multiple symmetries, the constant number of branches $\beta$, or the constant weight factor $\alpha$ of inhomogeneous networks. Many ``real" networks might not present these characteristics and, from the experimental point of view, there might be practical difficulties to reproduce them in a laboratory. However, the present study establishes a robust baseline to study more elaborated models where some of these restrictions could be relaxed. Current work is being carried out in that direction, with emphasis on computations of equivalent values in damaged networks, \ie networks with missing branches. In addition, the symmetries of self-similar trees allow the study of ``networks of networks'' (arrangements in which every single branch is another network),  which is a modern research field in network theory \cite{Agostino14}.

\end{document}